\documentclass[epj]{svjour}

\usepackage{colordvi}
\usepackage[T1]{fontenc}
\usepackage[english]{babel}
\usepackage{graphicx}

\newcommand{\be}{\begin{equation}}
\newcommand{\ee}{\end{equation}}
\newcommand{\bea}{\begin{eqnarray}}
\newcommand{\eea}{\end{eqnarray}}
\newcommand{\dd}{\mbox{d}} 
\newcommand\nn{\nonumber}

\begin{document}

\title{Processes $e^+e^-\to\pi^0({\pi^0}')\gamma$ in the NJL model}

\author{A.B. Arbuzov\inst{1,2} \and
E.A. Kuraev\inst{1} \and M.K. Volkov\inst{1}}

\institute{Bogoliubov Laboratory of Theoretical Physics,
JINR, Dubna, 141980  Russia \and
Department of Higher Mathematics, University Dubna,
Dubna, 141980   Russia}


\abstract{
The processes of electron-positron annihilation into $\pi^0\gamma$
and into $\pi'(1300)\gamma$ are considered within the NJL model.
Intermediate vector mesons $\rho^0$, $\omega$, $\rho'(1450)$,
and $\omega'(1420)$ are taken into account. The latter two mesons
are treated as the first radial excited states. They are incorporated
into the NJL model by means of a polynomial form factor. Numerical
predictions for the cross sections of these processes are received
for the center-of-mass energies below 2~GeV.
Our results for the $\pi^0\gamma$ production are in agreement with
experimental data received in the energy region 600 -- 1020 MeV.
}

\maketitle

\section{Introduction} \label{Intro}

In our recent work~\cite{Arbuzov:2010xi} the process $e^+e^-\to\pi^0\omega$
was considered in the framework of the extended Nambu--Jona-Lasinio (NJL)
model~\cite{Volkov:1996br,Volkov:1996fk,Volkov:1999yi}. In that work we took
into account $\rho(770)$ and $\rho'(1450)$ intermediate vector meson states.
The obtained results were found to be in a satisfactory agreement with the
experimental data~\cite{Akhmetshin:2003ag} at energies below 2~GeV.
Our results are also in a qualitative agreement with
the phenomenological description of this process
received within a vector meson dominance model~\cite{Akhmetshin:2003ag}
by fitting several free parameters from the experimental data.
It is worth to note that in our approach the process was described without
introduction of any additional arbitrary parameter. This testifies about the predictive
power of the NJL model and gives us a hope to obtain reliable predictions
for other similar processes in the same energy range.

In the present paper we consider the processes
$e^+e^-\to\pi^0({\pi^0}')\gamma$ which have small cross sections because
they are suppressed by factor $\alpha$ with respect to
other typical channels of electron-positron annihilation into hadrons.
Nevertheless, the process $e^+e^-\to\pi^0\gamma$ was studied with a
rather high accuracy in the energy regions around
the $\omega$ and $\phi$ meson masses \cite{Dolinsky:1991vq,Achasov:1990at,Achasov:2003ed}.
The ongoing high-luminosity experiments
in Novosibirsk (VEPP-2000) and Beijing (BES-III)
will also collect considerable statistics
for many possible annihilation processes including the ones with
production of $\pi'(1300)$.
So theoretical predictions for the given processes should be of interest
for the physical programs of these colliders.

\section{Lagrangian and process amplitudes} \label{Sec:1}

\begin{figure}
\resizebox{0.7\hsize}{!}{\includegraphics*{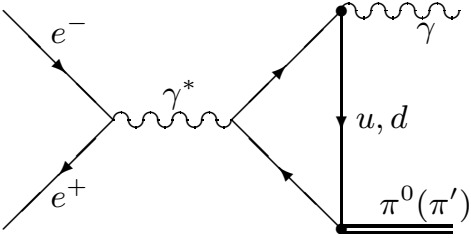}}
\caption{The Feynman diagram with photon exchange.
\label{Fig:1}}
\end{figure}

\begin{figure}
\resizebox{0.7\hsize}{!}{\includegraphics*{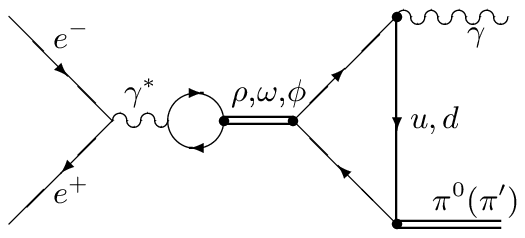}}
\caption{Feynman diagram(s) with $\rho^0$, $\omega$ and  $\phi$ meson exchange.
\label{Fig:2}}
\end{figure}

For the description of the first three diagrams (with intermediate $\gamma$, $\rho$, and $\omega$ states)
for the process of $\pi^0\gamma$ production,
see Figs.~\ref{Fig:1} and \ref{Fig:2},
we need the part of the standard NJL Lagrangian which describes interactions
of photons, pions, and the ground states of vector mesons with quarks,
see refs.~\cite{Volkov:1986zb,Volkov:1993jw,Ebert:1994mf}.
It has the form
\bea \label{L1}
&& \Delta{\mathcal L}_{1} = \bar{q}\biggl[i\hat\partial - m
+ \frac{e}{2}\biggl(\tau_3+\frac{1}{3}I\biggr)\hat{A}
\nonumber \\ && \quad
+ig_{\pi}\gamma_5\tau_3\pi^0
+ \frac{g_\rho}{2}\gamma_\mu\left(I\hat{\omega} + \tau_3\hat{\rho}^0\right) \biggr]q,
\eea
where $\bar{q}=(\bar{u},\bar{d})$ with $u$ and $d$ quark fields;
$m=diag(m_u,m_d)$, $m_u=m_d=280$~MeV is the constituent quark mass;
$e$ is the electron charge;
$A$, $\pi^0$, $\omega$ and $\rho^0$ are the photon, pion, $\omega$ and $\rho$ meson fields, respectively;
$g_\pi$ is the pion coupling constant,
$g_\pi=m_u/f_\pi$, where $f_\pi=93$~MeV is the pion decay constant;
$g_\rho$ is the vector meson coupling constant, $g_\rho\approx 6.14$ corresponding
to the standard relation $g_\rho^2/(4\pi)\approx 3$;
$I=diag(1,1)$ and $\tau_3$ is the third Pauli matrix.

\begin{figure}
\resizebox{0.7\hsize}{!}{\includegraphics*{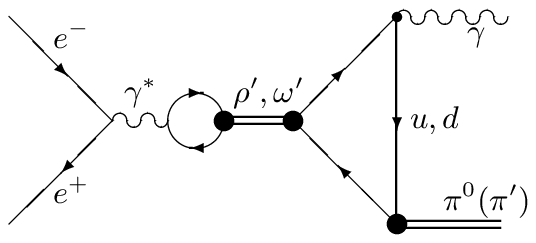}}
\caption{Feynman diagram(s) with $\rho'$ and $\omega'$ meson exchange.
\label{Fig:3}}
\end{figure}

For description of the radial excited mesons interactions we use the extended version
of the NJL Lagrangian~\cite{Volkov:1996br,Volkov:1996fk,Volkov:1997dd}:
\bea \label{L2}
&& \Delta {\mathcal L}_2^{\mathrm{int}} =
\bar{q}(k')\biggl\{ A_\pi \tau^3\gamma_5\pi^0(p)
- A_{\pi'} \tau^3\gamma_5\pi'(p)
\nn \\ && \quad
+ A_{\omega,\rho} \left(\tau^3{\hat{\rho}}(p)+I\hat{\omega}(p)\right)
 \\ \nn && \quad
- A_{\omega',\rho'} \left(\tau^3{\hat{\rho}'}(p)+I{\hat{\omega}'}(p)\right)
\biggr\}q(k),
\ \ \ p = k-k',
 \\ \nn
&& A_\pi = g_{\pi_1}\frac{\sin(\alpha+\alpha_0)}{\sin(2\alpha_0)}
       +g_{\pi_2}f({k^\bot}^2)\frac{\sin(\alpha-\alpha_0)}{\sin(2\alpha_0)},
\nn \\
&& A_{\pi'} = g_{\pi_1}\frac{\cos(\alpha+\alpha_0)}{\sin(2\alpha_0)}
       +g_{\pi_2}f({k^\bot}^2)\frac{\cos(\alpha-\alpha_0)}{\sin(2\alpha_0)},
\nn \\
&& A_{\omega,\rho} = g_{\rho_1}\frac{\sin(\beta+\beta_0)}{\sin(2\beta_0)}
       +g_{\rho_2}f({k^\bot}^2)\frac{\sin(\beta-\beta_0)}{\sin(2\beta_0)},
\nn \\ \nn
&& A_{\omega',\rho'} = g_{\rho_1}\frac{\cos(\beta+\beta_0)}{\sin(2\beta_0)}
        +g_{\rho_2}f({k^\bot}^2)\frac{\cos(\beta-\beta_0)}{\sin(2\beta_0)}.
\eea
The radially-excited states were introduced in the NJL model with the help of
the form factor in the quark-meson interaction:
\bea
f({k^\bot}^2) &=& (1-d |{k^\bot}^2|) \Theta(\Lambda^2-|{k^\bot}^2|),
\nn \\
{k^\bot} &=& k - \frac{(kp)p}{p^2},\qquad d = 1.78\ {\mathrm{GeV}}^{-2},
\eea
where $k$ and $p$ are the quark and meson momenta, respectively.
The cut-off parameter $\Lambda=1.03$~GeV is taken~\cite{Ebert:1992ag}.
The filled circles in fig.~\ref{Fig:3} denote the presence of the form
factor in the quark--meson vertexes.
Coupling constants $g_{\pi_1}=g_\pi$ and $g_{\rho_1}=g_\rho$ are the same as in
the standard NJL version. Constants $g_{\pi_2}=3.20$, $g_{\rho_2}=9.87$, and the mixing
angles $\alpha_0=59.06^\circ$, $\alpha=59.38^\circ$, $\beta_0=61.53^\circ$, and $\beta=76.78^\circ$
were defined in refs.~\cite{Volkov:1997dd,Arbuzov:2010vq}.

{\bf A. The process $e^++e^-\to\pi^0+\gamma$} contains contributions of three amplitudes:
\bea\label{Tlambda}
T^{\lambda} = \bar{e}\gamma_\mu e
\varepsilon_{\mu\lambda\alpha\beta}\frac{p_\pi^\alpha p_\gamma^\beta}{m s}
\bigl\{B_{\gamma}+B_{\rho+\omega+\phi}+B_{\rho'+\omega'}\bigr\},
\eea
where $s=(p_1(e^+) + p_2(e^-))^2$.
Contribution $A_{\gamma}^{\mu\lambda}$ corresponds to triangular diagram in fig.~{\ref{Fig:1}}, {\it i.e.}
the pion transition form factor:
\bea
B_{\gamma} = 2 V_{\gamma^*\pi^0\gamma}(s).
\eea
The sum of $\rho$ and $\omega$ meson contributions (see fig.~{\ref{Fig:2}}) reads
\bea
&& B_{\rho+\omega+\phi} = \biggl\{\frac{s}{s-M_\rho^2+iM_\rho\Gamma_\rho}
+\frac{s}{s-M_\omega^2+iM_\omega\Gamma_\omega}
\nn \\ && \quad
+\frac{s\sqrt{2}\sin\theta_{\omega\phi}}{s-M_{\phi}^2+iM_{\phi}\Gamma_{\phi}}
\biggr\}
\frac{1}{g_{\rho_1}} V_{\rho\pi^0\gamma}(s),
\eea
where $\gamma-\rho(\omega,\phi)$ transitions via quark loops~\cite{Volkov:1986zb}
are taken into account.
Note that in the case of $\omega$ meson the relative factor $1/3$ in the $\gamma-\omega$ transition
(with respect to the $\rho$ meson case) is canceled out with factor $3$ in the
$\omega\pi^0\gamma$ vertex.
Factor $\sqrt{2}$ ~in the numerator of the $\phi$ meson propagator arise from
the $\gamma-\phi$ through the $s$ quark loop.
The standard value of the $\phi-\omega$ mixing angle $\theta_{\omega\phi}\approx -3^\circ$
is used~\cite{Volkov:1993jw,Ebert:1994mf,Gronau:2009mp}.

Let us emphasize that the sum of the photon, $\rho$, and $\omega$ meson contributions gives
the expression coinciding with the result of vector meson dominance (VMD) model, which
emerges from the NJL model, see also~\cite{Volkov:1986zb,Volkov:2006vq,Volkov:1982zx,Ebert:1982pk}.

The contributions of the excited mesons are calculated in an analogous
way but with the extended Lagrangian~(\ref{L2}).
We get
\bea
&& B_{\rho'+\omega'} =
\biggl( - \frac{\cos(\beta+\beta_0)}{\sin(2\beta_0)}
- \Gamma \frac{\cos(\beta-\beta_0)}{\sin(2\beta_0)} \biggr)
\frac{1}{g_{\rho_1}}
\\ \nn && \quad \times
\biggl\{\frac{s}{s-M_{\rho'}^2+iM_{\rho'}\Gamma_{\rho'}}
+\frac{s}{s-M_{\omega'}^2+iM_{\omega'}\Gamma_{\omega'}}
\biggr\}
\\ \nn && \quad \times
 V_{\rho'\pi^0\gamma}(s).
\eea

We checked that the possible effect of taking into account the running 
of the meson widths is small. So for the numerical calculations we use the
values from the Particle Data Group~\cite{Nakamura:2010zzi}: 
$\Gamma_\rho=146.2$~MeV,
$\Gamma_\omega=8.49$~MeV,
$\Gamma_{\rho'}=400$~MeV, and
$\Gamma_{\omega'} = 215$~MeV.

The vertexes are defined via the triangular loop integrals of the anomalous type:
\bea
&& V_{\gamma^*\pi^0\gamma} = g_{\pi_1} I^{(3)}_0,
\\ \nn
&& I^{(3)}_n = - \int\frac{\dd^4k\; m^2 f^n({k^\bot}^2)
 \Theta(\Lambda^2-|{k^\bot}^2|)} {i\pi^2\; (k^2-m^2+i0)}
\\ \nn
&&\quad \times
\frac{1}{((k+p_\gamma)^2-m^2+i0)((k-p_\pi)^2-m^2+i0)},
\eea

\bea \label{loop_int}
&& V_{\rho\pi^0\gamma} = g_{\pi_1}
\biggl(
 \frac{\sin(\beta+\beta_0)g_{\rho_1}I^{(3)}_0}{\sin(2\beta_0)}
+\frac{\sin(\beta-\beta_0)g_{\rho_2}I^{(3)}_1}{\sin(2\beta_0)}
\biggr),
\nonumber \\ \nonumber
&& V_{\rho'\pi^0\gamma} = -g_{\pi_1}
\biggl(
 \frac{\cos(\beta+\beta_0)g_{\rho_1}I^{(3)}_0}{\sin(2\beta_0)}
+\frac{\cos(\beta-\beta_0)g_{\rho_2}I^{(3)}_1}{\sin(2\beta_0)}
\biggr),
\eea
The vertex factors for $\omega$ mesons $V_{\omega\pi^0\gamma}$
and $V_{\omega'\pi^0\gamma}$ are just three times greater than
the $\rho$ meson ones.
In the $p^2$ approximation the triangular diagrams reproduce the`<
Wess-Zumino terms in the chiral symmetric meson 
Lagrangian~\cite{Volkov:1986zb,Volkov:1993jw,Ebert:1994mf,Ebert:1985kz}.
With help of these terms one can describe radiative decays of all pseudoscalar and vector meson
nonets in a satisfactory agreement with experimental data~\cite{Volkov:1986zb,Volkov:1993jw,Ebert:1994mf}.
In this work we will use the same approximation\footnote{The cut-off in the loop integral~(\ref{loop_int})
will be used only in the presence of the form factor.}.


The transition of photon into $\rho$ meson reads~\cite{Volkov:1986zb}:
\bea \label{gamma_rho}
\frac{e}{g_\rho} \bigl( g^{\nu\nu'}q^2 - q^{\nu}q^{\nu'}\bigr).
\eea
The $\gamma-\omega$ transition differs from the above just by factor $1/3$.
In the amplitudes with excited mesons, we have to take into account
the $\gamma-\rho_2$ and $\gamma-\omega_2$  transitions
 ($\gamma-\rho_1(\omega_1)$ ones are the same as in the
standard $\gamma-\rho(\omega)$ cases)
can be expressed via the $\gamma-\rho(\omega)$
transition with the additional factor~\cite{Volkov:1996fk,Volkov:1997dd}
\bea
\Gamma = \frac{I_2^f}{\sqrt{I_2I_2^{f^2}}}\approx 0.47.
\eea
In particular, the $\gamma-\rho'$ transition takes the form
\bea \nn
\frac{e}{g_\rho} \bigl( g^{\nu\nu'}q^2 - q^{\nu}q^{\nu'}\bigr)\biggl\{
\frac{\sin(\beta+\beta_0)}{\sin(2\beta_0)} + \Gamma \frac{\sin(\beta-\beta_0)}{\sin(2\beta_0)}
\biggr\}.
\eea
Note that the relative factors $1/3$ (in photon-meson
transitions) and $3$ (in vertexes)
in contributions of $rho$ and $\omega$ mesons cancel each other.

{\bf B. The process $e^++e^-\to\pi'(1300)+\gamma$} can be described is a very similar manner to
the previous one. The main difference is in the vertexes $V_{\gamma^*(\rho,\rho',\omega,\omega')\pi'\gamma}$.
They read
\bea
&& V_{\gamma^*\pi'\gamma} = V_{\gamma^*\pi^0\gamma} \frac{\cos(\alpha+\alpha_0)}{\sin(2\alpha_0)}
+ g_{\pi_2}I^{(3)}_1\frac{\cos(\alpha-\alpha_0)}{\sin(2\alpha_0)},
\nn \\ \nn
&& V_{\rho\pi'\gamma} = - g_{\rho_1} \frac{\sin(\beta+\beta_0)}{\sin(2\beta_0)}
g_{\pi_1}\frac{\cos(\alpha+\alpha_0)}{\sin(2\alpha_0)}I^{(3)}_0
\\ \nn && \quad
- g_{\rho_2}\frac{\sin(\beta-\beta_0)}{\sin(2\beta_0)}
g_{\pi_1}\frac{\cos(\alpha+\alpha_0)}{\sin(2\alpha_0)}I^{(3)}_1
\\ \nn && \quad
-  g_{\rho_1}\frac{\sin(\beta+\beta_0)}{\sin(2\beta_0)}
g_{\pi_2}\frac{\cos(\alpha-\alpha_0)}{\sin(2\alpha_0)}I^{(3)}_1
\\ \nn && \quad
-  g_{\rho_2}\frac{\sin(\beta-\beta_0)}{\sin(2\beta_0)}
g_{\pi_2}\frac{\cos(\alpha-\alpha_0)}{\sin(2\alpha_0)}I^{(3)}_2,
\nn \\ \nn
&& V_{\rho'\pi'\gamma} =  g_{\rho_1} \frac{\cos(\beta+\beta_0)}{\sin(2\beta_0)}
g_{\pi_1}\frac{\cos(\alpha+\alpha_0)}{\sin(2\alpha_0)}I^{(3)}_0
\\ \nn && \quad
+ g_{\rho_2}\frac{\cos(\beta-\beta_0)}{\sin(2\beta_0)}
g_{\pi_1}\frac{\cos(\alpha+\alpha_0)}{\sin(2\alpha_0)}I^{(3)}_1
\\ \nn && \quad
+  g_{\rho_1}\frac{\cos(\beta+\beta_0)}{\sin(2\beta_0)}
g_{\pi_2}\frac{\cos(\alpha-\alpha_0)}{\sin(2\alpha_0)}I^{(3)}_1
\nn \\ && \quad
+  g_{\rho_2}\frac{\cos(\beta-\beta_0)}{\sin(2\beta_0)}
g_{\pi_2}\frac{\cos(\alpha-\alpha_0)}{\sin(2\alpha_0)}I^{(3)}_2.
\eea
All other factors are the same as in the pion production case.
But the propagators of $\rho'$ and $\omega'$ mesons can be taken
with constant widths: $\Gamma_{\rho'}=340$~MeV and
$\Gamma_{\omega'}=215$~MeV.

Note that within the same approach the listed here
vertexes allow us to get a good  description
the radiative decays of meson ground states:
$\Gamma_{\pi^0\to\gamma\gamma}=7.7$~eV,
$\Gamma_{\rho^0\to\pi^0\gamma}=77$~keV,
$\Gamma_{\omega\to\pi^0\gamma}=710$~keV.
The corresponding experimental values~\cite{Nakamura:2010zzi} are
$\Gamma^{\mathrm{exp}}_{\pi^0\to\gamma\gamma}=7.8\pm 0.5$~eV,
$\Gamma^{\mathrm{exp}}_{\rho^0\to\pi^0\gamma}=88\pm 12$~keV,
$\Gamma^{\mathrm{exp}}_{\omega\to\pi^0\gamma}=700\pm 30$~keV.

\section{Cross section and Numerical Results}

Now we can estimate the contributions of the considered amplitudes into
the total process cross section. The details of phase volume calculations
and evaluation of the cross section can be found in
ref.~\cite{Bystritskiy:2009zz}. For our case it takes the form
\bea \label{sigma}
&& \sigma^{e^+e^-\to\pi\gamma}(s) =
\frac{\alpha^3}{24\pi^2s^3f_\pi^2}\lambda^{3/2}(s,M_\omega^2,M_\pi^2)
\frac{1}{g_{\pi_1}^2}
\nn \\ && \quad \times
|B_{\gamma}+B_{\rho+\omega+\phi}+B_{\rho'+\omega'}|^2
\\ \nn
&& \lambda(s,M_\omega^2,M_\pi^2) = (s-M_\omega^2-M_\pi^2)^2-4M_{\omega}^2M_{\pi}^2.
\eea
The cross section for $\pi'\gamma$ production has the same form with simple substitutions.

\begin{figure}
\resizebox{0.9\hsize}{!}{\includegraphics*{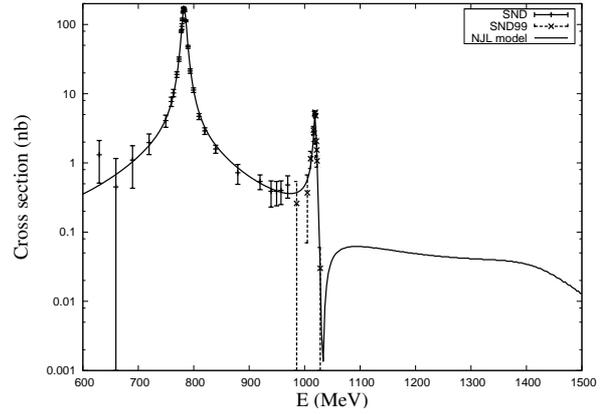}}
\caption{Comparison of experimental results for $e^+e^-\to\pi^0\gamma$
with the NJL model prediction.
\label{plot}}
\end{figure}

Fig.~\ref{plot} shows the experimental data of
the SND collaboration~\cite{Achasov:1990at,Achasov:2000zd}
and the corresponding theoretical result (the solid line)
received within the applied here NJL phenomenological model. 
The NJL model predictions are in a good agreement with the experimental
data. In particular, we found the following the values for the cross 
section at the $\omega$ and $\phi$ peaks:
\bea
&& \sigma^{e^+e^-\to\pi\gamma}({m_\omega}^2) = 177~{\mathrm{nb}}, 
\nonumber \\ 
&& \sigma^{e^+e^-\to\pi\gamma}({m_\phi}^2) = 5.5~{\mathrm{nb}}. 
\eea  

\begin{figure}
\resizebox{0.9\hsize}{!}{\includegraphics*{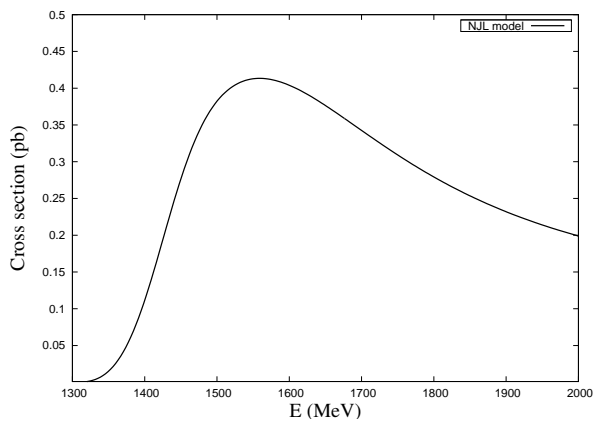}}
\caption{NJL model prediction for the cross section of $e^+e^-\to\pi'\gamma$ process.
\label{pipr_gamma}}
\end{figure}

Predictions for the value of the cross section of annihilation into the $\pi(1300)\gamma$
pair are given in fig.~\ref{pipr_gamma}. The peak due to $\rho'$ and $\omega'$ meson resonances
is clearly seen there, but the magnitude of the peak is small. Partially that is due to 
mutual compensation of the ground end excited states in $V_{\rho'\pi'\gamma}$. 
Note that the NJL model is adjusted for applications at low energies up to about 2~GeV.
In this energy range,
the model gives a qualitative description of meson properties and interactions. The
advantage is that the set of parameters is limited and fixed.
It worth to stress that to describe the given processes we did
not introduce any new parameter in the model.

\section{Conclusions}

Our calculation for the process $e^+e^-\to\pi^0\gamma$ showed the presence of two resonance
regions in the energy range below 1.1~GeV. The first resonance appears in the region of the
$\omega$ meson mass and look as a very high narrow peak.
The second one is also narrow, it lies in the region of
$\phi$ meson mass. The deep just after the $\phi$ meson peak is due to destructive interference
of amplitudes.
At the present time this process
is studied with high accuracy only in the energy region below 1~GeV~\cite{Achasov:2003ed}.
It worth to note that the experimental energy dependence of the process cross section in the first
both resonance region is very close to our theoretical estimation. 
Ongoing experiments at VEPP-2000
and BES-III collect statistics at higher energies. This will allow us to perform a comparison
of our theoretical predictions with future experimental data. The cross section of $\pi'\gamma$
production has the order of about 1~pb, so it is difficult to measure.

Further we are going to consider radiative processes with participation of
$\eta$, $\eta'$, $\rho$, $\omega$, $\phi$ mesons and their radial excited states
in the framework of the extended $U(3)\times U(3)$ NJL model. We plan to consider
also production of these particles in $e^+e^-$ collisions.

\begin{acknowledgement}
We are grateful to A.~Akhmedov for useful discussions.
This work was supported by RFBR grant 10-02-01295-a.
\end{acknowledgement}

\end{document}